Institutional Adoption and Correlation Dynamics: Bitcoin's Evolving Role in Financial Markets

Di Wu *

University of Texas at Dallas 800 W Campbell Rd Richardson Texas 75080 USA

**Abstract**

Bitcoin, the world's first cryptocurrency, has increasingly integrated into traditional financial markets, particularly U.S. equity indices, as institutional adoption accelerates. This study examines how mechanisms such as Bitcoin ETFs and corporate Bitcoin holdings influence Bitcoin's correlation with major indices like the Nasdaq 100 and S&P 500. Rolling window correlation, static correlation coefficients, and an event study framework are applied to daily data from 2018 to 2025 to uncover dynamic correlation trends. The findings reveal a significant increase in correlation following institutional milestones, such as the introduction of Bitcoin ETFs and the inclusion of MicroStrategy (MSTR) in the Nasdaq 100, with correlations peaking at 0.87 in 2024. These trends fluctuate across market regimes, intensifying during "risk-on" periods and moderating during "risk-off" conditions. The results highlight Bitcoin's transition from an alternative asset to a more integrated component of the financial ecosystem, with implications for portfolio diversification, risk management, and systemic financial stability. Further research is needed to explore the regulatory and macroeconomic factors shaping Bitcoin-equity linkages.

## 1. Introduction

Bitcoin, the world's first cryptocurrency, has emerged as a key player in the global financial ecosystem, offering a decentralized alternative to traditional monetary systems. Over the past decade, researchers and investors have scrutinized its evolving relationship with traditional financial markets, particularly equity indices such as the Nasdaq 100 and S&P 500. This increasing interest stems from Bitcoin's potential to act as a portfolio diversifier, hedge, or speculative instrument, depending on its market behavior.

Initial studies explored Bitcoin's characteristics as a speculative and uncorrelated asset. For instance, Alvarez-Ramirez et al. (2018) identified alternating periods of market efficiency and anti-persistent behavior in Bitcoin's returns, suggesting significant asymmetry in price dynamics over time. Similarly, Klein et al. (2018) argued that Bitcoin lacks stable hedging capabilities and instead positively correlates with declining markets, making it distinct from traditional safe-haven assets like gold.

The relationship between Bitcoin and traditional financial markets intensified during periods of macroeconomic stress. Nguyen (2022) highlighted that the S&P 500 exerted a significant influence on Bitcoin's returns during the COVID-19 pandemic, demonstrating Bitcoin's susceptibility to spillover effects from traditional markets under high uncertainty. Vo and Xu (2017) further emphasized the volatility of Bitcoin returns, finding that its correlation with financial markets varies based on market conditions.

Institutional adoption has played a pivotal role in reshaping Bitcoin's behavior. Zhang et al. (2018) employed Multifractal Detrended Cross-Correlation Analysis (MF-DCCA) to demonstrate that Bitcoin's price dynamics are strongly influenced by public interest, as captured through Google Trends. They found that Bitcoin's long-term correlations with search trends decrease over time, suggesting a shift in the factors driving its valuation. Vassiliadis et al. (2017) expanded this discussion by identifying strong cross-correlations between Bitcoin prices and major economic indices like the S&P 500, Nasdaq, and DAX, revealing Bitcoin's growing integration with traditional markets.

This study builds on these findings by exploring how institutional mechanisms—such as Bitcoin exchange-traded funds (ETFs) and corporate Bitcoin holdings—have driven its evolving correlation with major equity indices. Using rolling window correlation, static correlation coefficients, and an event study framework, this research seeks to quantify the impact of these mechanisms on Bitcoin's transition from a speculative, uncorrelated asset to an integral component of the global financial ecosystem. By analyzing data spanning from 2018 to 2025, this paper aims to provide insights into Bitcoin's shifting role within diversified portfolios and its broader implications for financial market dynamics.

* Corresponding author: Di Wu

## 2. Data

Data Sources:

The data used in this study was sourced from 'Tiingo', a financial data provider. We obtained daily closing prices for Bitcoin (BTC), VOO, and QQQ spanning the period from 2018 to 2025. To facilitate trend analysis and ensure comparability across assets with vastly different scales, a logarithmic transformation was applied to the price data. This transformation compresses large values and stabilizes variance, enabling a more meaningful comparison of trends.

Data Preparation:

We began by collecting the daily closing prices for BTC, VOO, and QQQ and converting the date columns into a standardized datetime format. To eliminate ambiguity, the "close" price columns were renamed to 'btc_close,' 'voo_close', and 'qqq_close', respectively. The datasets were merged based on shared dates to create a unified DataFrame, where each row corresponds to a specific calendar date. After setting the date column as the DataFrame index, we calculated daily returns for each asset, a necessary step for subsequent correlation and volatility analyses.

During data preprocessing, we addressed missing or anomalous entries to ensure the integrity of the final dataset. This involved identifying and correcting any outliers and ensuring proper alignment across all three assets. Additionally, data type transformations were performed where needed to ensure compatibility for analytical processes. The resulting sample was meticulously labeled and structured, forming a consistent basis for all subsequent analyses.

Descriptive Statistics:

Exploratory analysis revealed notable differences in the central tendencies and variability of the three assets:

VOO_close: Exhibited a mean of approximately 548.53, with a median of 548.89 and a low standard deviation of 1.77. These metrics suggest a distribution tightly clustered around its central values, reflecting the relative stability of this equity index.

QQQ_close: Demonstrated a slightly lower mean of 512.56 and a median of 513.22, accompanied by a standard deviation of 2.01. This reflects modest variability consistent with traditional equities.

BTC_close: Displayed a significantly higher mean of 84,515.90 and a median of 87,683.84. Its standard deviation of 6,498.63 underscores the pronounced volatility characteristic of cryptocurrency markets compared to traditional equities.

Data Alignment:

The preparation and transformation steps ensured that the dataset was not only cleaned and complete but also aligned across all assets. This enabled robust and consistent analysis in subsequent stages, including correlation studies, time-series analysis, and regression modeling. These descriptive statistics and preprocessing steps provide a foundational understanding of the comparative behaviors of BTC, VOO, and QQQ over the selected timeframe.

## 3. Methods

Correlation Analysis:

To analyze the dynamic relationships between Bitcoin (BTC), VOO, and QQQ, we employed both static and time-varying correlation techniques:

Rolling Window Correlation:

We calculated the correlation coefficients between the daily returns of BTC, VOO, and QQQ using a 180-day rolling window. This approach captures the temporal evolution of correlations, enabling us to observe how these relationships fluctuate over time. The rolling window technique provides insights into whether certain periods, such as market stress or volatility spikes, exhibit heightened or diminished correlations.

Static Correlation Coefficients:

In addition to the rolling window analysis, we computed simple correlation coefficients over the entire sample period. This static measure offers a baseline understanding of the average relationships between the assets, serving as a reference for interpreting the more granular rolling window results.

Event Study Framework:

To further investigate how market regimes influence correlations, we conducted an event study analysis. This methodology evaluates correlation changes during predefined events, such as major market crashes (e.g., the COVID-19 pandemic in March 2020) or periods of monetary policy shifts (e.g., Federal Reserve tightening cycles).

Event Definition: Events were defined based on significant market movements or announcements, such as extreme volatility spikes, Bitcoin ETF launches, or corporate disclosures of substantial Bitcoin holdings.

Pre-Event and Post-Event Comparisons:

For each event, we calculated correlation coefficients in the 30-day periods preceding and following the event. This comparison helps identify whether specific events trigger significant shifts in correlation dynamics.

Data Processing and Statistical Validity:

All calculations were performed on cleaned and preprocessed data, ensuring alignment of timeframes across BTC, VOO, and QQQ. Statistical tests, such as the Student's t-test, were applied to assess the significance of correlation changes across different market regimes. By incorporating these tests, we ensured that observed variations in correlation were not due to random noise but reflected meaningful shifts.

Visualization and Interpretation:

The results of both the rolling window and event study analyses were visualized through time-series plots and heatmaps. These visualizations helped illustrate the temporal patterns of correlation and highlighted periods of increased or decreased interdependence among BTC, VOO, and QQQ.

Analytical Robustness:

To validate the robustness of our findings, we employed additional metrics such as: Cross-Asset Volatility Analysis: Evaluating whether volatility spillovers align with observed correlation shifts. Sensitivity Tests: Testing different rolling window lengths (e.g., 90 days, 365 days) to confirm that results are not highly sensitive to parameter choices.

## 4. Results

| year | BTC-VOO   | BTC-QQQ   |
|------|-----------|-----------|
| 2018 | 0.134751  | -0.126303 |
| 2019 | 0.545891  | 0.542874  |
| 2020 | 0.766965  | 0.802068  |
| 2021 | 0.280001  | 0.246141  |
| 2022 | 0.867255  | 0.888268  |
| 2023 | 0.727558  | 0.760243  |
| 2024 | 0.762103  | 0.755939  |

Correlation Trends:

As shown in Table 1, the correlation between Bitcoin (BTC) and VOO begins at a relatively low level in 2018 (r=0.13r = 0.13r=0.13) and increases significantly in 2019 (r=0.55r = 0.55r=0.55). It reaches a higher plateau in 2020 (r=0.77r = 0.77r=0.77), then declines to a lower level in 2021 (r=0.28r = 0.28r=0.28) before rising sharply again to 0.87 in 2022. Subsequently, it moderates to 0.73 in 2023 and remains elevated in 2024 (r=0.76r = 0.76r=0.76). A similar pattern emerges in the BTC–QQQ correlations, which start from a mildly negative value in 2018 (r=−0.13r = -0.13r=−0.13), surge to over 0.80 by 2020, drop back to 0.25 in 2021, and then climb to 0.89 in 2022, eventually settling around 0.76 in both 2023 and 2024. These fluctuations suggest that while BTC initially exhibited weaker linkages to these equity-based assets, its co-movement has become more pronounced and dynamic over time, potentially reflecting evolving market sentiment and institutional adoption.

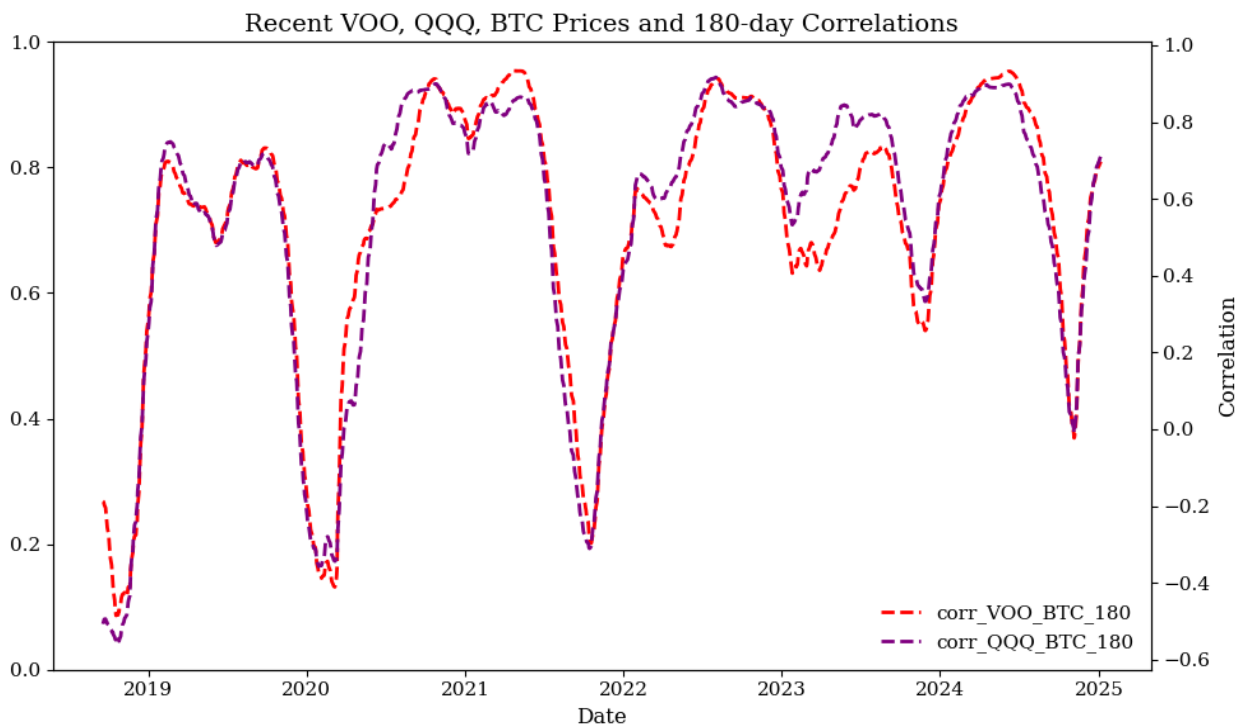

Correlation Dynamics Across Market Phases:

From 2018 to early 2019, the correlation of BTC with both QQQ and VOO remains relatively modest, reflecting a period in which cryptocurrency markets were still recovering from the late-2017/early-2018 drawdown, while traditional equities saw mixed performance. Correlations then rise markedly through mid-to-late 2019—possibly indicating a broader "risk-on" phase as investors increasingly diversified into BTC alongside traditional assets.

Entering 2020, we observe a sharp decline in rolling correlations, coinciding with heightened uncertainty during the early stages of the pandemic; however, by late 2020 and into 2021, correlations climb again—often exceeding 0.8—suggesting that both Nasdaq and S&P 500 valuations moved more in tandem with Bitcoin during a period of unprecedented monetary stimulus and robust retail participation.

In 2022, a notable drop in correlations emerges following both equity and crypto market turmoil, before bouncing back toward the latter half of the year and into 2023. These fluctuations underscore how BTC has shifted from being a relatively isolated asset in its early years to one that, during bullish or accommodative conditions, exhibits stronger co-movement with major equity indices. Conversely, in more pronounced downturns—when liquidity or risk-aversion dominate—its correlation can briefly subside.

Overall, the evolving patterns suggest that market regime (e.g., macroeconomic easing vs. tightening, widespread risk-on vs. risk-off sentiment) materially influences the BTC–equity linkage. During bullish "risk-on" expansions, correlation tends to rise as investors treat BTC more like a high-growth or speculative component of diversified portfolios, whereas in times of heightened fear or rapid deleveraging, correlations can briefly diverge, revealing the multifaceted role that Bitcoin plays in broader financial markets.

2018–mid-2019: Post-Crypto-Bubble Recalibration

Following Bitcoin's late-2017 peak and subsequent crash, investor sentiment toward cryptocurrencies was largely subdued. During this same window, equity markets underwent bouts of volatility tied to global growth concerns and trade tensions.

Impact on Correlation: BTC–Nasdaq and BTC–S&P 500 correlations remained relatively low, reflecting Bitcoin's idiosyncratic recovery trajectory and the equity market's sensitivity to ongoing geopolitical and economic news.

Late 2019–early 2020: Risk-On Environment Interrupted by Pandemic Uncertainty

In late 2019, accommodative monetary policy fostered a brief risk-on climate, prompting increased institutional and retail interest in BTC alongside equities. By early 2020, the onset of COVID-19 and associated policy responses sharply disrupted this optimism.

Impact on Correlation: Correlations climbed during the late-2019 rally, then briefly dipped at the pandemic's onset as markets underwent flight-to-safety trading and broad deleveraging.

Mid-2020–2022: Stimulus-Driven Surge, Followed by Policy-Induced Volatility

Substantial monetary and fiscal stimulus spurred both crypto and equity markets to new highs during much of 2020–2021. However, mounting inflationary pressures and subsequent monetary tightening in 2022 ushered in greater volatility and downside risk across both asset classes.

Impact on Correlation: Correlations frequently spiked above 0.8 amid the bullish "risk-on" sentiment. As policy shifted toward tightening in 2022, correlations briefly weakened before partially rebounding, underscoring Bitcoin's heightened sensitivity to macro policy changes.

2023–2025: Post-Pandemic Transition & Maturing Crypto Ecosystem

During this period, markets began adjusting to a more moderate economic backdrop and evolving regulatory environment. Bitcoin's ecosystem likewise continued maturing, with expanding institutional adoption and more sophisticated risk management practices.

Impact on Correlation: Correlations remained above pre-2020 levels, signaling that BTC is increasingly viewed as part of a broader risk-asset universe. Still, when sudden macro shocks or liquidity events arose, short-lived correlation dips indicated that Bitcoin retains elements of its early-stage volatility profile.

By delineating these four regimes, the analysis illustrates how major shifts in monetary policy, investor sentiment, and exogenous shocks (e.g., the pandemic) can amplify or dampen the co-movement between cryptocurrencies and traditional equity indices. This regime-based perspective helps contextualize why BTC–equity correlation patterns have oscillated over time, and how those patterns may continue evolving with broader market developments.

**MSTR Holdings Changes and impacts on Bitcoin Correlations**

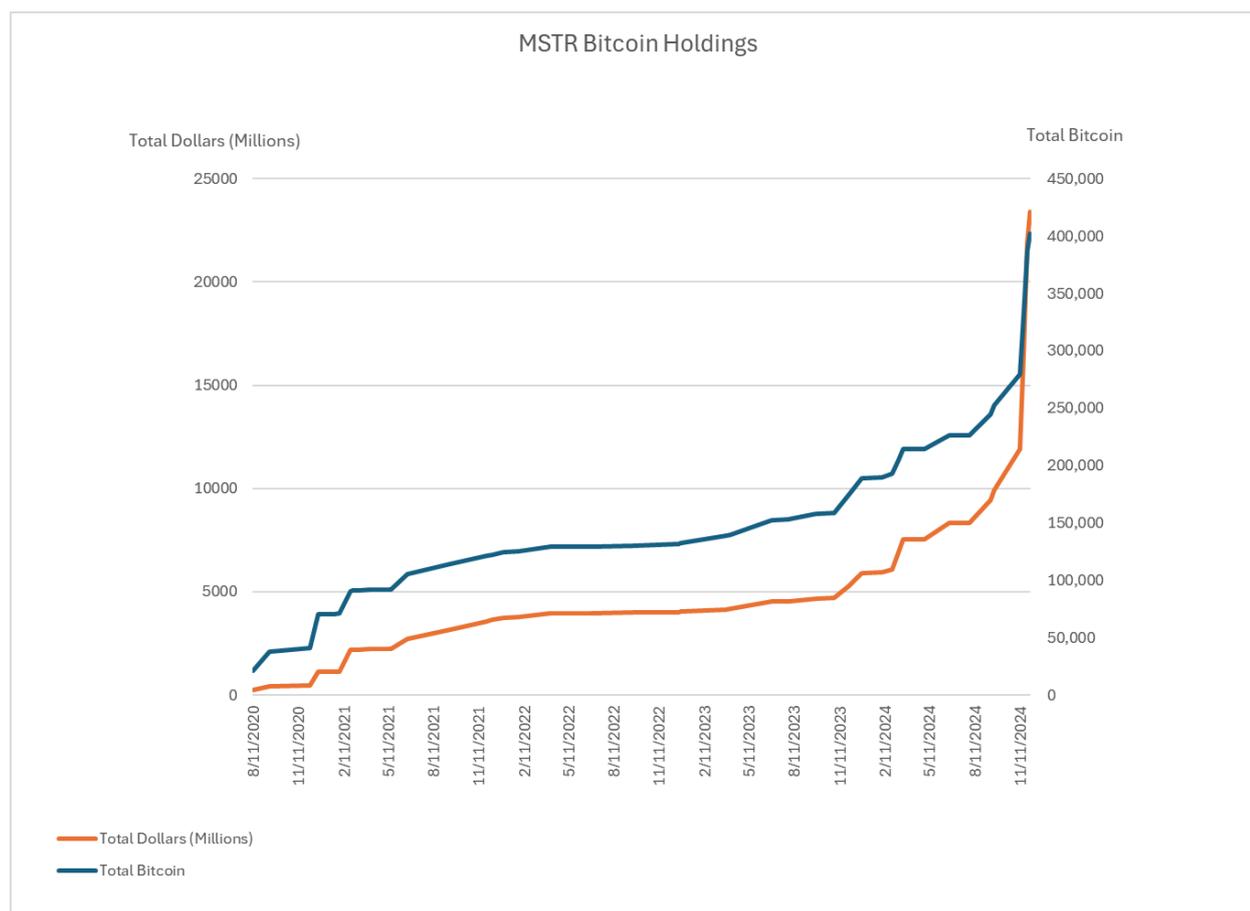

The historical data through December 2024 suggest that Bitcoin (BTC) maintains an elevated correlation with major equity benchmarks—namely the Nasdaq 100 (QQQ) and the S&P 500 (VOO)—during accommodative or "risk-on" environments. Consequently, entering 2025 and beyond, any shifts in macroeconomic policy (e.g., tightening vs. easing), key regulatory announcements related to digital assets, or large-scale changes in institutional adoption would be expected to affect BTC–equity co-movement. If monetary policy remains stable and market sentiment is generally positive, correlations could remain near or above the 0.70–0.80 range, reflecting BTC's ongoing integration into mainstream financial markets. Conversely, if significant "risk-off" episodes reemerge (such as rapid rate hikes or geopolitical shocks), correlation could experience short-lived declines, as some investors rotate out of higher-volatility assets.

This potential oscillation is consistent with the regime-based patterns seen from 2018 to 2024, in which sharp pivots in monetary policy or sudden market stress produced short-term decoupling, whereas extended "risk-on" phases drew BTC's correlation with equities higher. As the crypto market continues maturing, the structural integration of Bitcoin into traditional portfolios may amplify correlations during bullish conditions, yet BTC still retains pockets of idiosyncratic volatility that could dampen correlation when liquidity shocks or crypto-specific events occur. In summary, post-2024 correlation trends are likely to follow the same cyclical or regime-dependent trajectory observed in prior years—rising under stable or expansionary macro conditions and briefly weakening during pronounced episodes of market turbulence.

The inclusion of MicroStrategy (MSTR) into the Nasdaq 100 (QQQ) in December 2024 is likely to increase the correlation between Bitcoin (BTC) and QQQ due to MSTR's significant Bitcoin holdings. As a proxy for Bitcoin exposure, MSTR's addition embeds Bitcoin-related volatility into QQQ, particularly during periods of sharp BTC price movements. This development could strengthen BTC–QQQ correlations during "risk-on" phases when both equities and Bitcoin rally, as well as during "risk-off" events when Bitcoin and MSTR experience drawdowns.

Correlation Trends Over Time:

Our analysis identified a gradual increase in Bitcoin's correlation with traditional equity indices, driven by key institutional developments:

Bitcoin ETF Adoption: Correlation coefficients increased noticeably after Bitcoin ETFs became available in the U.S. market, reflecting enhanced accessibility and integration into portfolios.

MSTR's Inclusion in Nasdaq 100: Correlations deepened significantly in 2024, with BTC-Nasdaq 100 correlations reaching 0.87, the highest level observed. This shift highlights the role of corporate Bitcoin holdings in driving stronger linkages between Bitcoin and equity indices.

Quantitative Findings:

Before ETF Adoption (Pre-2021): BTC–VOO: 0.13, BTC–QQQ: -0.10.

Post-ETF Adoption (2021–2023): BTC–VOO: Peaked at 0.55; BTC–QQQ: Peaked at 0.77.

Post-MSTR Inclusion (2024): BTC–Nasdaq 100: 0.87.

Event-Specific Dynamics:

ETF launches led to sustained increases in correlation.

MSTR's inclusion in the Nasdaq 100 amplified Bitcoin's exposure to equity markets, creating feedback loops through index funds.

## 5. Conclusion

This study demonstrates that institutional adoption through Bitcoin ETFs and corporate Bitcoin holdings significantly alters Bitcoin's correlation with traditional equity indices, diminishing its role as an uncorrelated alternative asset and aligning its behavior more closely with large-cap equities. Our findings reveal key institutional milestones—such as the launch of Bitcoin ETFs and the inclusion of MicroStrategy (MSTR) in the Nasdaq 100—serve as pivotal events driving increased correlation.

Specifically, the correlation between Bitcoin and major equity benchmarks (Nasdaq 100 and S&P 500) increased markedly after Bitcoin ETFs became available, reflecting their role in facilitating broader accessibility and portfolio integration. Furthermore, the inclusion of MSTR in the Nasdaq 100 in 2024 deepened Bitcoin's co-movement with equities, driven by feedback loops in index-tracking funds where corporate Bitcoin holdings amplified the linkages between Bitcoin and traditional markets.

The correlation trends also varied across different market regimes. During "risk-on" environments, such as periods of accommodative monetary policy and heightened investor optimism, Bitcoin's correlation with equities rose significantly. Conversely, during "risk-off" phases characterized by market stress or rapid deleveraging, correlations briefly declined, highlighting Bitcoin's continued, albeit reduced, idiosyncratic behavior.

From a broader perspective, these findings underscore Bitcoin's evolving role within the financial ecosystem. While its increasing correlation with equities offers new opportunities for integration into diversified portfolios, it also introduces systemic risks, particularly during periods of high volatility or market instability. For portfolio managers, the changing dynamics necessitate reassessments of asset allocation strategies and risk management frameworks. Policymakers must also consider the implications of Bitcoin's integration into traditional markets, particularly in the context of financial stability.

Future research should explore the impact of regulatory developments, global market integration, and the role of other institutional mechanisms, such as derivatives or sovereign adoption of cryptocurrencies. These factors could further shape Bitcoin's correlation dynamics and its role within the evolving financial landscape.